# Second-Harmonic Generation of Spoof Surface Plasmon Polaritons Using Nonlinear Plasmonic Metamaterials

Hao Chi Zhang[1], Yifeng Fan[1], Jian Guo[1], Xiaojian Fu[1], Lianming Li[1], Cheng Qian[1], and Tie Jun Cui[1,2†]

[1] *State Key Laboratory of Millimeter Waves, Southeast University, Nanjing 210096, China*

[2] *Cooperative Innovation Centre of Terahertz Science, No.4, Section 2, North Jianshe Road, Chengdu 610054, China*

† E-mail: tjcui@seu.edu.cn

**The second harmonic generation is one of the most important applications of nonlinear effect, which has attracted great interests in nonlinear optics and microwave in the past decades. To the best of our knowledge, however, generating the second harmonics of surface plasmon polaritons (SPPs) has not been reported. Here, we propose to generate the second harmonics of spoof SPPs with high efficiency at microwave frequencies using subwavelength-scale nonlinear active device integrated on specially designed plasmonic waveguides, which are composed of two ultrathin corrugated metallic strips printed on the top and bottom surfaces of a thin dielectric slab anti-symmetrically. We show that the plasmonic waveguide supports broadband propagations of spoof SPPs with strong subwavelength effect, whose dispersion property can be controlled by changing the geometrical parameters. By loading the nonlinear device made from semiconductors to the intersection of two plasmonic waveguides with different corrugation depths, we experimentally demonstrate the efficient generation of second-harmonic SPPs in broad frequency band. The proposed second-harmonic generator can be directly used as SPP frequency multiplier, and the proposed method can be extended to achieve high-order harmonics and produce SPP mixers, which are essential to SPP integrated circuits and systems.**

At optical frequencies, the natural surface plasmon polaritons (SPPs) are special surface electromagnetic (EM) waves, which crawl along the interface between metal and dielectric materials. SPPs have intrigued great interests in physics during the past decades due to the potential for developing new types of photonic devices [1, 2]. The interactions between free electrons of conductor and EM fields around give rise to unique properties of SPPs, such as the significant field confinements in the vicinity of the metal-dielectric interface and great enhancements of EM energies [3-5]. Motivated by the recent advances that allow metals to be structured and characterized on nanometer scales, SPPs have found promising applications in the super-resolution imaging [6,7], miniaturized sensors [8], and photovoltaics [9], etc. The SPP-based optical circuits have been considered as a solid venue for the future developments [10].

However, the method of structuring SPPs cannot be used in the microwave and terahertz frequencies because metals fundamentally exhibit as perfect electric conductors below the far infrared regime, instead of plasmas with the negative permittivity. In order to achieve highly confined SPPs at lower frequencies, plasmonic metamaterials have been proposed [11-20], whose performances can be designed at will through changing the geometrical parameters of surface structures. Spoof SPPs have inherited the most exotic features of optical SPPs, such as the field confinement and non-diffraction limit. Recently, an ultrathin corrugated metallic strip has been proposed to propagate conformal surface plasmons (CSPs) on arbitrarily curved surfaces [21], which represents one of the most potential candidates for applications of SPP devices and circuits in microwave and terahertz frequencies owing to the ultrathin structure, low loss, broadband, and flexibility.

The strong EM fields resulting from SPPs allow weakly nonlinear process, which depends linearly on the local fields [22]. In previous studies, nonlinear media have been employed to enhance the nonlinear effects [23-29], such as nonlinear crystals in the optical regime and split-ring resonators with packaged varactors embedded in capacitive gaps in the microwave regime [23]. Since electrically-large and bulky nonlinear media are required, it is difficult to realize compact and efficient SPP nonlinear devices. More importantly, the required large intensity of incident waves to excite the nonlinear phenomena is much stronger than that used in applications (e.g. communications and sensing), which makes it difficult to realize high-efficiency nonlinear SPP devices.

As one of the most important applications of nonlinear effects, the second harmonic generation has attracted great interests in nonlinear optics and microwave in the past decades. However, to the best of our knowledge, the second-harmonic generation of SPP waves has not been reported. Most of previous studies involving nonlinear surface plasmons were aimed to control spatial lights or EM waves by the enhanced nonlinear effects [30-32]. In this work, we propose to realize the second harmonics of spoof SPPs at microwave frequencies. Taking the advantage of semiconductor technology in the microwave regime, we propose to use nonlinear active chip in subwavelength scale to produce high-order harmonics of spoof SPPs, as illustrated in **Fig. 1(a)**. Our purpose is not only to verify the physical phenomena of nonlinear SPPs, but also to realize SPP frequency multiplier and mixer with high gain and high efficiency for real applications. To integrate the subwavelength-scale nonlinear chip into spoof SPP structures conveniently, we use a specially designed plasmonic waveguide [33], which is constructed by printing two narrow corrugated metallic strips on the top and bottom surfaces of a thin dielectric substrate with mirror symmetry. The numerical simulations and experimental results demonstrate the efficient generation of second-harmonic SPP waves propagating along the plasmonic waveguides.

Comparing to the ultrathin corrugated metallic strip [21], the specially designed plasmonic waveguide has many superiorities and its dispersion performance can be designed at will by tuning the geometrical parameters [34]. We optimize the geometries of the input and output plasmonic waveguides so as to cover the working bandwidth of the nonlinear active chip, and the optimized geometrical parameters are presented in Table 1. The corresponding dispersion curves of the input and output structures are illustrated in **Fig. 2**. Owing to the easy tune of wave momentum, we design a smooth conversion between the SPP waves and the conventional guided waves [35]. The conversion part functions as the momentum matching and impedance matching between the traditional microstrip line and plasmonic waveguides.

In the past, nonlinear metamaterials have been proposed [23], in which a packaged varactor was embedded in the metamaterials to enhance the nonlinear effect. However, a powerful incident field was required, and hence the nonlinear phenomenon is hard to observe and difficult to be applied in engineering.

To solve the problem, we propose to employ a field effect transistor (FET), which is a kind of active audion, to generate the second harmonics of spoof SPPs. The nonlinear effect of FET is more significant than that of packaged varactor, which makes it easier to generate the high-order harmonics. More importantly, the output characteristics of FET can be tuned by the direct current (DC) bias, and this strategy is easily implemented by the semiconductor technology. If we use a single FET to generate the second-harmonic SPPs, the conversion gain is calculated analytically as

$$G = V_{2-order} / V_{input} = 2\sum_{n=1}^{\infty} \frac{q^{(2n)}(V)}{(2n!!)^2} V_{input}^{2n} \frac{n}{n+1} \qquad (1)$$

in which, $V_{input}$ is the amplitude of the input SPP signal, $V$ is the DC bias, and $q(\bullet)$ represents the voltage response function of FET.

To achieve significant conversion gain and restrain other higher-order harmonics, we apply a simple FET model in our design, as illustrated in **Fig. 3(a)**, which consists of a differential amplifier circuit that acts as an active balun. There is an 180° phase difference between two outputs of the balun. Then the outputs feed the gates of balanced FETs and the drains are connected to form the single-ended output, which results in the cancellation of fundamental frequency and odd harmonics, and the in-phase superposition of even-harmonic drain currents. In the figure, the node 'S' represents a virtual ground.

For simplicity, we analyze this strategy using the circuit method. We assume that the input signal can be expressed as a function of angular frequency, while the DC bias remains the same in the process. The final output of the two FETs is calculated analytically as

$$V_{out} = q(V+V_{in}(\omega)) + q(V-V_{in}(\omega)) = \sum_{n=1}^{\infty} \frac{2q^{(2n)}(V)}{2n!} V_{in}^{2n}(\omega) \qquad (2)$$

in which $V_{in}(\omega)$ is the input SPP signal as a function of angular frequency. Since high-order terms exist, the nonlinear field is generated through the combination of two FETs. To obtain a certain harmonic, we can change the DC bias ($V$) to make the FET work in an appropriate voltage range. Except FETs, the whole system can be regarded as a linear system. According to the dispersion spectrum of spoof SPPs, the transmission spectrum of the SPP waveguide corresponds to a low-pass filter, which can effectively suppress the higher-order harmonics. The dispersion property of spoof SPPs, which can be engineered at will by changing the geometrical parameters, provides greater convenience to design functional devices than that of traditional transmission lines.

Hence we use two transmission-spectrum functions $A_1(\omega)$ and $A_2(\omega)$ to describe the response of the linear input part (including the input spoof SPP waveguide) and differential amplifier, and the response of the linear output part (including the output SPP waveguide), respectively. For easy understanding, we consider the simplest case, i.e., the single-frequency incidence. Then the conversion gains from the fundamental frequency to the second harmonic is calculated analytically as

$$G = V_{2-order}/V_{input} = 4A_2(\omega) \sum_{n=1}^{\infty} \left( A_1^{2n}(\omega) V_{input}^{2n-1} \frac{q^{(2n)}(V)}{(2n!!)^2} \frac{n}{n+1} \right) \tag{3}$$

which indicate that the second harmonic is significantly enhanced.

To verify the generation of second SPP harmonics at microwave frequencies, we consider a realistic FET device in numerical simulations. Here, we choose a commercially available AMMC-6120 nonlinear active chip, which is commonly used in the conventional active microwave devices and circuits, and the kernel is adaptive to our strategy. The other parts of AMMC-6120 can be regarded as a matching network to the 50 ohm. We apply the AMMC-6120 active chip to produce the nonlinear second harmonics and control the input power at 0 dBm. From the previous discussion, the SPP waveguide is an ideal filter. Hence the conversion gain is estimated as 11 dB from 5 to 10 GHz according to Eq. (3).

For quantitative analysis, we perform full-wave simulations using the CST Microwave Studio. The whole structure is illustrated in **Fig. 1(a)**, in which metal and substrate materials are selected as copper and F4B ($\varepsilon_r = 2.65$, $\tan\delta = 0.001$). The geometrical parameters of input and output plasmonic waveguides are listed in Table 1. We connect the active chip to the plasmonic waveguide efficiently [34] and define the input and output of the active chip as Port 2 and Port 3, respectively, as shown in **Fig. 3(b)**. The simulated S-parameters of the structure are illustrated in **Fig. 4(a)**, in which S41 shows that almost no energy flows to the output port directly, S21 indicates that most of the input energy transfers to the input of the active chip through the input plasmonic waveguide, while S43 implies that most of the second-harmonic energy transfers to the output port through the output plasmonic waveguide.

For direct observation of the field distribution in the whole structure, we make Ports 1 and 3 excite signals simultaneously. **Fig. 5** demonstrates the electric-field distributions on the structure surface when the fundamental frequency is 8 and 10 GHz, which are jointed by the electric fields of input waveguide at the fundamental frequency and output waveguide at the second-harmonic frequency. We clearly observe that the fundamental SPP modes on the input waveguide are converted to the second-harmonic SPP modes on the output waveguide efficiently, with significant amplifications of the converted fields. We also notice that both fundamental and second-harmonic SPP fields are tightly confined in deep sub-wavelength scales around the waveguides with low propagation loss.

Based on the printed circuit board (PCB) technology, we fabricate a sample of active plasmonic waveguide with the same parameters used in the aforementioned simulations, whose bottom and top views are illustrated in **Figs. 3(c)** and (**d**), respectively. The spectrum measured results are presented in Table 2. As a typical example, the measured frequency spectrum at 8 GHz is demonstrated in **Fig. 4(b)**, from which the second harmonic is clearly observed at 16 GHz with significant gain, accompanying a very weak third harmonic at 24 GHz.

From Table 2, we note that the measured conversion gain from the fundamental frequency to the second harmonic is around 10 dB when the fundamental frequency changes from 5 to 10 GHz, which is very close to the theoretical prediction 11dB. Hence the second harmonic SPPs are efficiently generated in a wide frequency band under the current design. From the same table, we also observe small measured conversion gain from the fundamental frequency to the third harmonic, which is caused by the cut-off characteristic of the output plasmonic waveguide. The significant reduction of the third harmonic is one of distinctive advantages of SPPs over the traditional transmission lines. The small discrepancies between theoretical analysis and measured results may originate from the loss of SPP waveguide, the assemble error, and bonding wire effect in experiments, which have not been considered in the previous analysis.

To show the second-harmonic SPP waves in experiments, we measure the near electric fields on an observation plane at 1.5 mm above the active SPP structure by an improved near-electric-field mapper [34]. The measured near-field distributions are illustrated in **Fig. 6**, in which (**a**) and (**b**) illustrate the electric fields at the fundamental frequency of 8 GHz and second harmonic of 16 GHz, while (**c**) and (**d**) present the electric fields at fundamental frequency of 10 GHz and second harmonic of 20 GHz. In both cases, we clearly observe the generations of second-harmonic SPPs through a subwavelength-scale active chip from the measured near fields, which propagate very well to the output port.

In summary, we proposed to generate the second-harmonic spoof SPPs with significant gains using a subwavelength-scale nonlinear chip in broadband microwave frequencies. To incorporate the nonlinear chip efficiently, we adopted two corrugated metallic strips that are printed on the top and bottom surfaces of a dielectric substrate anti-symmetrically. Through the nonlinear active chip, we experimentally realized the second-harmonic generation of SPPs, in which the measured conversion gains are around 10 dB from 6 to 10 GHz, having good agreements to the theoretical analysis. We also directly measured the second-harmonic SPPs from the near electric fields. Considering the high efficiency and excellent performance, the second harmonic generator can be used as an SPP frequency multiplier in real applications.

The proposed method can be directly extended to produce higher-order harmonics of SPPs in the microwave frequency, realizing high-performance SPP mixers. We may also use the method to generate SPP multipliers and mixers in the millimeter wave and terahertz frequencies. Together with other active and passive SPP components [33, 35-40], the presented work provides a promising basis to build up large-scale spoof SPP integrated circuits and systems in the microwave and terahertz regimes.

This work was supported by the National Science Foundation of China (61171024, 61171026 and 61138001), the National High Tech (863) Projects (2012AA030402 and 2011AA010202), and the 111 Project (111-2-05).

———————————

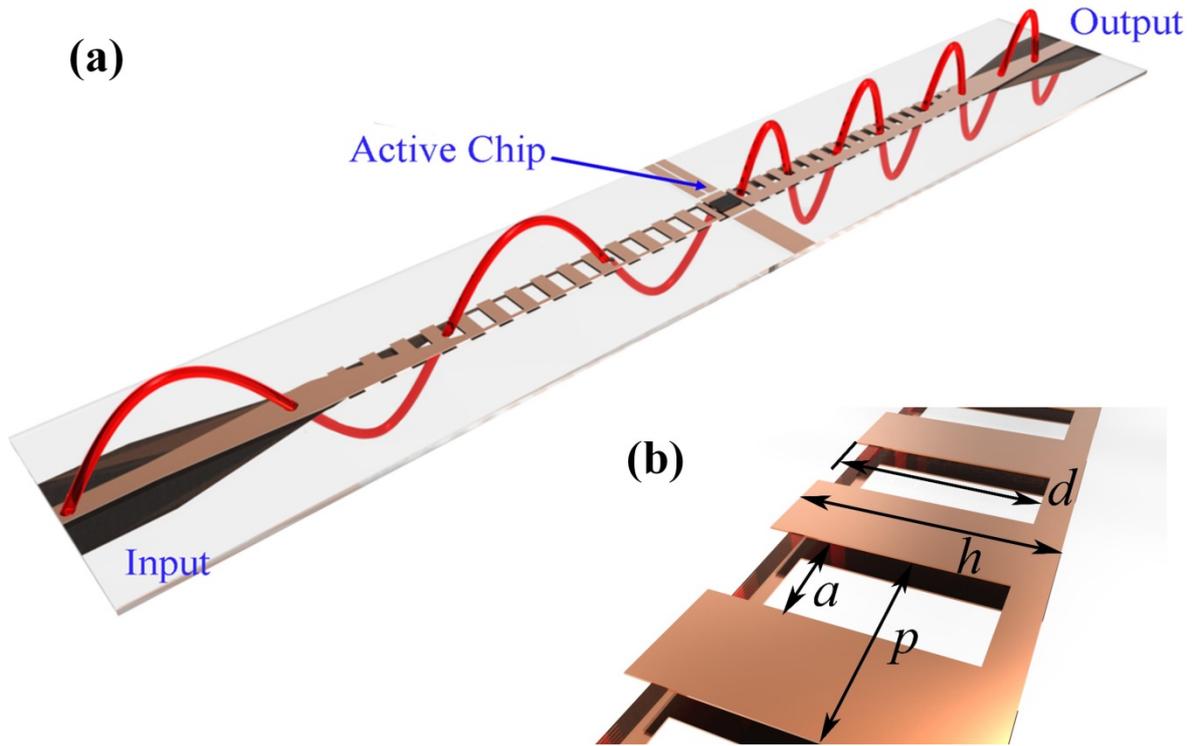

**FIG. 1** Second-harmonic generation of spoof SPPs. (a) Schematic diagram to generate the second-harmonic SPPs through a subwavelength-scale nonlinear active chip integrated in a special plasmonic waveguide composed of two anti-symmetrical metallic strips. (b) The detailed structure of the special plasmonic waveguide, in which the geometry configuration of a unit cell is described by the length $p$, groove depth $d$, groove width $a$, and strip width $w$.

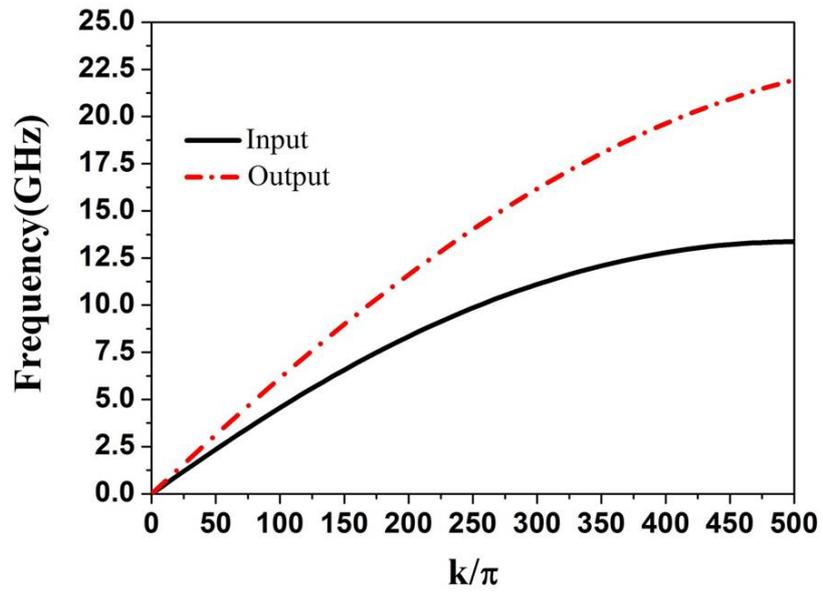

**FIG. 2** Dispersion diagrams of the input and output plasmonic structures.

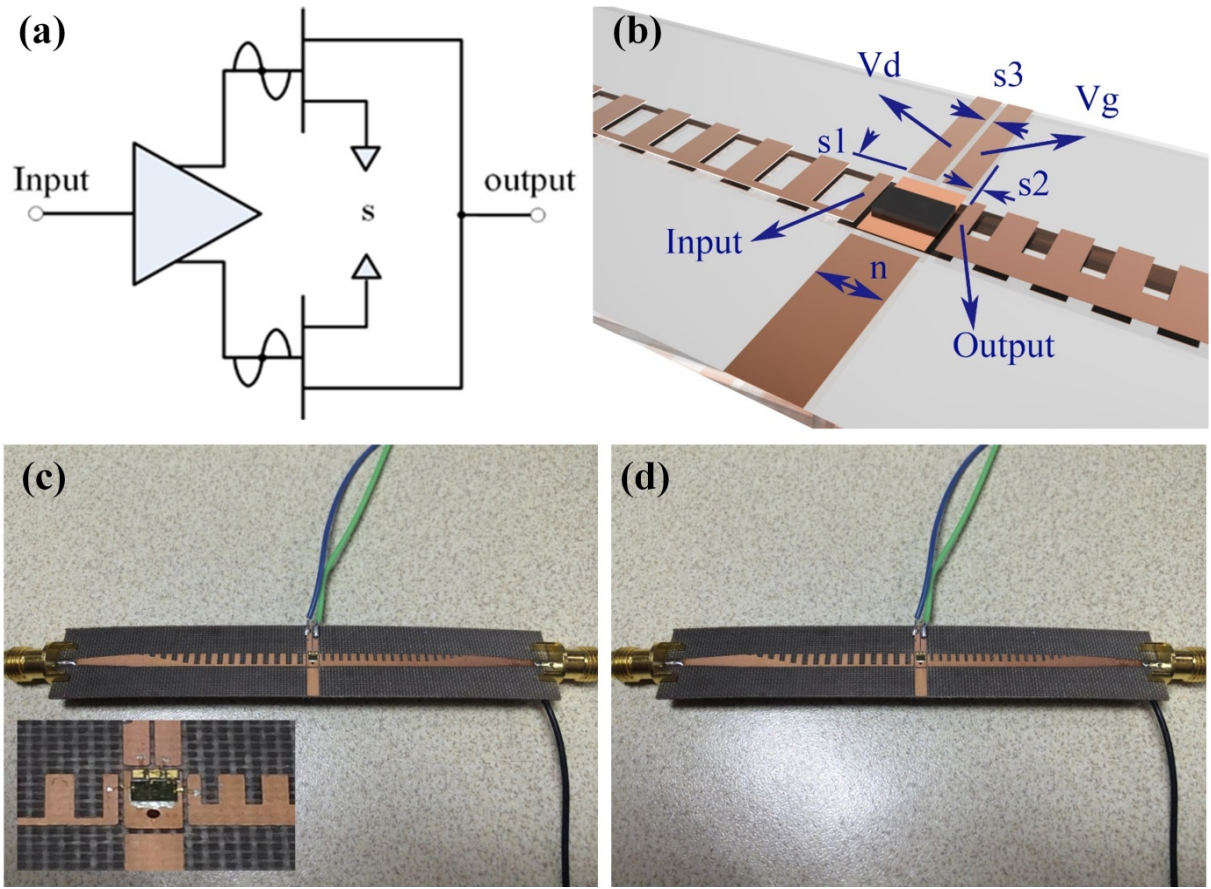

**FIG. 3** Design and fabrication of the second-harmonic spoof SPP generator. (a) The topology diagram of the proposed strategy to generate second-harmonic SPPs. (b) The top view of the installation section for nonlinear active chip, in which $n$ is the length of the central pad, and S1 and S2 are the gaps between the central pad and the structures around it. (c, d) Photographs of the fabricated sample (top view and bottom view) of the second-harmonic SPP generator. The inset of (c) shows the enlarged view of the active chip mounted on the plasmonic waveguide.

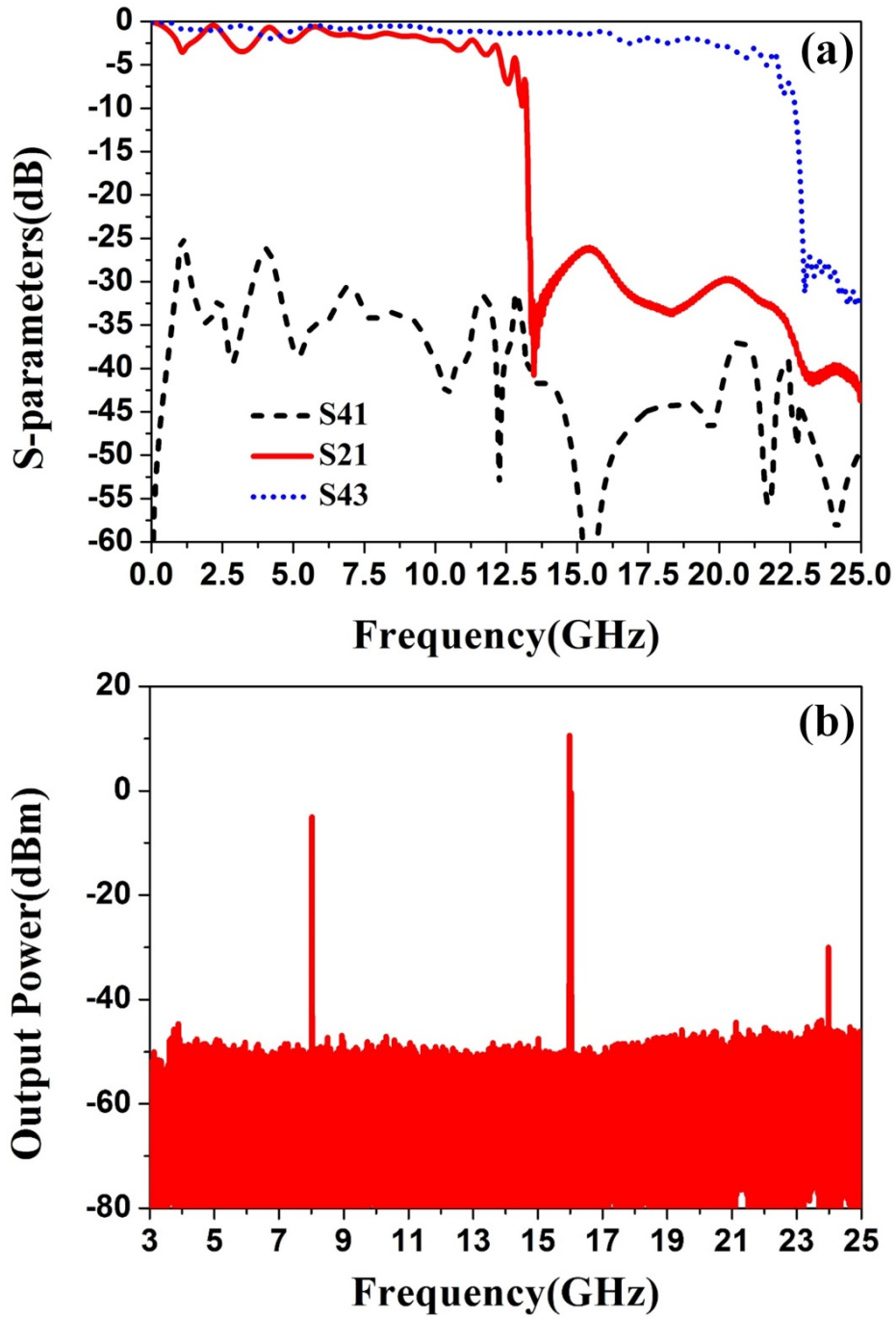

**FIG. 4** Simulation and measurement results of the second-harmonic SPP generator. (a) The simulated scattering S parameters of the whole structure, in which the input and output of the nonlinear active chip are replaced by two ports (3 and 4), in which the material types of metal and substrate are selected as copper and F4B. (b) The measurd frequency spectrum of the second-harmonic SPP generator when the fundamental frequency is 8 GHz.

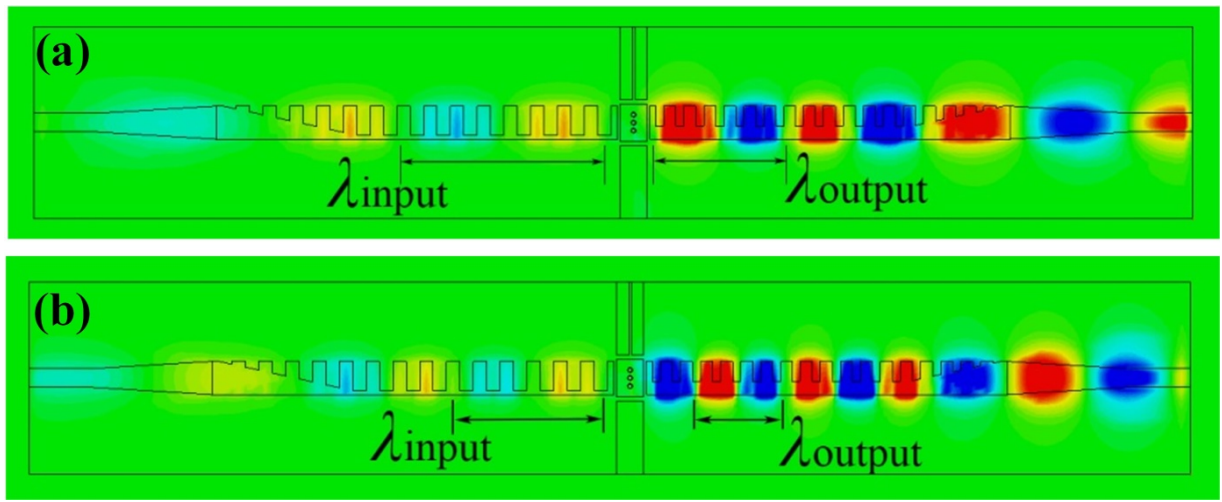

**FIG. 5** The simulated near-electric-field distributions on the plasmonic structure surface at different fundamental frequencies $f$, in which the black outline indicates the region of spoof SPP waveguides. (a) $f = 8$ GHz. (b) $f = 10$ GHz.

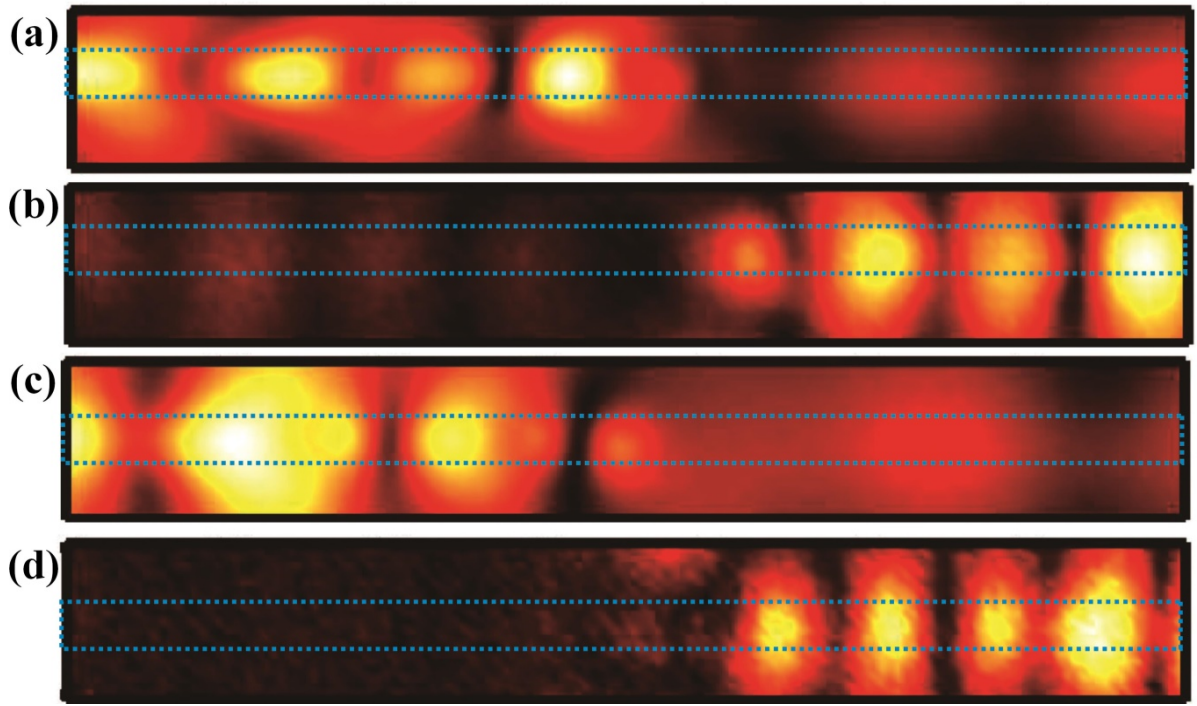

**FIG. 6** The measured near-electric-field distributions on the observation plane that is 1.5 mm above the plasmonic structure at different frequencies, in which the geometry parameters are shown in Table 1. (a) The electric fields monitored at 8 GHz (fundamental frequency) when the input frequency is 8 GHz. (b) The electric fields monitored at 16 GHz (second harmonic) when the input frequency is 8 GHz. (c) The electric fields monitored at 10 GHz (fundamental frequency) when the input frequency is 10 GHz. (d) The electric fields monitored at 20 GHz (second harmonic) when the input frequency is 10 GHz.

**Table 1** The geometrical parameters of the input and output plasmonic waveguides

|                  | Period *p* | Groove width *a* | Groove depth *d* |
|------------------|------------|------------------|------------------|
| Input waveguide  | 2.0 mm     | 1.0 mm           | 2.2 mm           |
| Output waveguide | 1.5 mm     | 0.6 mm           | 1.5 mm           |

**Table 2** Experimental results in generating the second-harmonic spoof SPPs

| Input Frequency (Input Power: 0dBm) | First Harmonic (dBm) | Second Harmonic (dBm) | Third Harmonic (dBm) |
|---|---|---|---|
| 5 GHz  | -15.0 | 11.1 | -3.6  |
| 6 GHz  | -17.0 | 10.0 | -6.2  |
| 7 GHz  | -16.0 | 9.6  | -7.8  |
| 8 GHz  | -5.0  | 10.6 | -30.0 |
| 9 GHz  | -2.8  | 10.0 | -48.0 |
| 10 GHz | -8.6  | 10.2 | -55.0 |

Supplemental Material

# Second-Harmonic Generation of Spoof Surface Plasmon Polaritons Using Nonlinear Plasmonic Metamaterials

Hao Chi Zhang[1], Yifeng Fan[1], Jian Guo[1], Xiaojian Fu[1], Lianming Li[1], Cheng Qian[1], and Tie Jun Cui[1, 2†]

[1] *State Key Laboratory of Millimeter Waves, Southeast University, Nanjing 210096, China*

[2] *Cooperative Innovation Centre of Terahertz Science, No.4, Section 2, North Jianshe Road, Chengdu 610054, China*

† E-mail: tjcui@seu.edu.cn

In this supplemental material, we provide detailed descriptions of the special plasmonic waveguide and experimental setups.

## 1. Features of the Special Plasmonic Waveguide

It has been shown that the ultrathin corrugated metallic strip is a good SPP waveguide in the microwave and terahertz frequencies [21]. However, it cannot be integrated with an active device owing to the single-line structure. To solve the problem, a special plasmonic waveguide has been proposed [33], which is composed of two anti-symmetrically corrugated metallic strips. The mirror structure is capable of connecting the nonlinear active device efficiently, besides exhibiting stronger sub-wavelength and field confinement effects. **Fig. 1(b)** depicts the unit cell of the plasmonic waveguide, in which the groove depth and width are denoted as *d* and *a*, and the strip width and thickness are *t* and *h*, respectively. The whole structure is constructed by arranging such unit cells along the *x* axis with period *p* and its mirror duplicate on the top and bottom surfaces of a dielectric substrate with thickness $t_s$. We remark that the substrate can be designed as ultrathin and flexible, if required.

It has been demonstrated that the equivalent plasmonic frequency of the mirror structure is much smaller than that of the single comb-shaped SPP waveguide [21] under the same geometrical configuration. We show that the mirror structure possesses more significant frequency dispersions than the single corrugated strip with the cutoff frequency decreasing rapidly. For specification, the dispersion curve of two anti-symmetrically corrugated metallic strips deviates from the light line more significantly. This unique feature results in smaller propagating wavelength and tighter electromagnetic field confinements of the proposed SPP waveguide under the same frequency, predicting promising applications such as the miniaturized SPP devices and low-interference transmission lines.

We consider two cases of the special plasmonic waveguide, which support spoof SPPs in two frequency bands. To reach the design, we investigate the dispersion properties of the mirror structures with different periods, groove depths, and widths. Based on the available metallic film and dielectric substrate, the geometrical parameters are chosen as *h*=2.5mm, *t*=0.018mm, and $t_s$=0.5mm. The metal is selected as copper, and the dielectric substrate is F4B with the dielectric constant $\varepsilon_r = 2.65$ and loss tangent $\tan\delta = 0.001$. We carry out the numerical simulations by the commercial software, CST Microwave Studio. **Fig. S1(a)** shows the dispersion curves of the mirror structure with the fixed period and groove width (*p*=2.0mm and *a*=1.0mm) and varying groove depth *d*. We observe that the curve with *d*=0, which corresponds to two parallel metal strips, is very similar to the dispersion curve of a traditional microstrip line and is well distinct to the light line. As $d$ increases from 0.8 to 2.4 mm, the dispersion curve deviates gradually from that of the microstrip and further from the light line, and asymptotically approaches to different cutoff frequencies. This is similar to the phenomenon originally appeared in the optical frequencies for natural SPPs.

We further investigate the influences of structure period and groove width on the wave momentum and cutoff frequency. When the period and groove depth are fixed ($p$=2.0 mm and $d$=2.4 mm) while the groove width $a$ varies, the dispersion relations are illustrated in **Fig. S1(b)**, in which the curve with $a$=0 again corresponds to two parallel metal strips. As $a$ increases from 0.3 to 1.2 mm, the dispersion curve does not exhibit so significant changes as observed in **Fig. S1(a)**, implying that the dispersion characteristic of the mirror structure is insensitive to the groove width. In this regard, the proposed plasmonic waveguide allows slightly tuning of wave momentum by adjusting the groove width. In **Fig. S1(c)**, the groove depth and width are fixed as $d$=2.4 mm and $a$=1.0 mm. When the period $p$ increases from 0.3 to 1.2 mm, the dispersion curve remains unchanged in the transmission band. However, the period influences the cutoff frequency slightly, and hence it is an effective approach to tune the cutoff frequency independently in small ranges by changing the period.

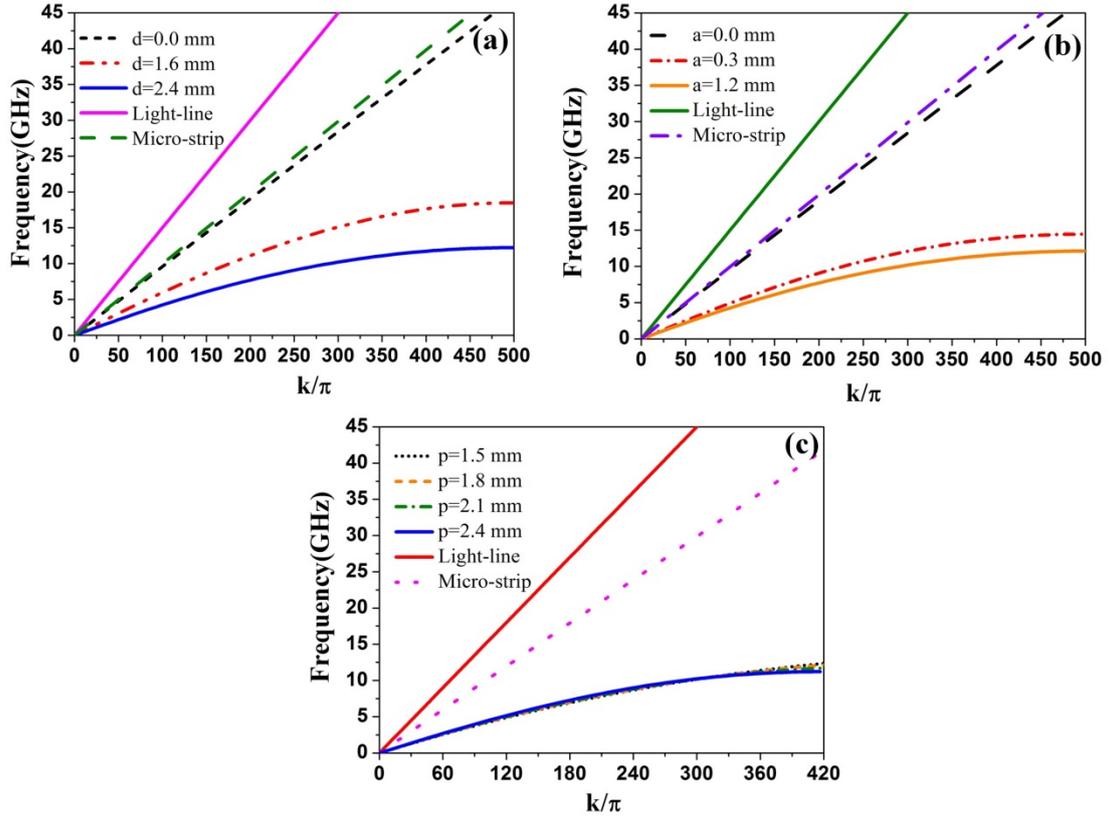

**FIG. S1** Dispersion diagrams of different plasmonic structures. (a) The plasmonic structures when the groove depth ranges from 0 to 2.4 mm. (b) Plasmonic structures when the groove width ranges from 0 to 1.2 mm. (c) Plasmonic structures when the period ranges from 1.5 to 2.4 mm. Other parameters in all scenarios are fixed as: $h$=2.5 mm, $t$=0.018 mm, and $t_s$=0.5 mm. The material types of metal and substrate are selected as copper and F4B.

## 2. Detailed Description in Connecting the Plasmonic Waveguide to the Nonlinear Active Chip

As shown in **Fig. 3(b)**, the input, output and DC bias pads of the nonlinear active chip are connected to points 'input', 'output', 'Vd' and 'Vg' on the double-side plasmonic structure, respectively. The central pad, at which the chip is located, is connected to the bottom metallic strip through metallic via holes. As a consequence, the chip and plasmonic waveguide share the same ground, which is essentially impossible for the single-side SPP waveguide [21]. All geometrical parameters of the structure around the active chip have been carefully optimized to guarantee good impedance matching, which enables the EM energy to bump into and flow out of the chip efficiently.

The commercially available AMMC-6120 active chip is attached onto the central pad using electrically conductive adhesive. Each welding spot is connected to the corresponding unit cell on the structure using bonding wire, as shown in Fig. 3(b).

## 3. The Measurement of Nonlinear Spectra

To guarantee the pure DC input, three bypass capacitors of 100 pF are added among the Vd and Vg pads on the plasmonic waveguide and the DC bias welding spot on the active chip. By welding two standard SMA connectors to the two ports of plasmonic waveguide, the frequency spectrum of the whole system is obtained by using the Agilent signal generator (E8257D) and Agilent spectrum analyzer (E4447A). The Agilent signal generator is connected to the input waveguide and the Agilent spectrum analyzer is connected to the output waveguide ($V_d$=4.5 V; $V_g$=1.6 V).

## 4. The Improved Near-Electric-Field Mapper

The traditional method of the near-field measurement system relies heavily on a vector network analyzer (VNA). But most VNAs are linear systems, which require the frequency of measured near fields be the same as the frequency of input signal. To overcome the problem, we propose a new near-field measurement system, which is composed of a planar platform, an Agilent VNA (N5230C), an Agilent signal generator (E8257D), a detector, and a motion controller. The fabricated active sample is put on the planar platform, which can move in the *x*- and *y*-directions automatically controlled by the motion controller. The input port of the SPP waveguide is connected to the signal generator through SMA to feed the EM energy, an electric-monopole detector is connected to one port of VNA to probe the vertical components of electric fields, and two broadband matching loads are connected to the other port of VNA and the output port of active SPP waveguide, respectively. Then we obtain the near-electric-field scanning on the measurement plane with a step resolution of 0.5 mm.